\documentclass[12pt]{article}
\newlength{\extraspace}
\newlength{\extraspaces}
\newcommand{\be}{\begin{equation}
\addtolength{\abovedisplayskip}{\extraspaces}
\addtolength{\belowdisplayskip}{\extraspaces}
\addtolength{\abovedisplayshortskip}{\extraspace}
\addtolength{\belowdisplayshortskip}{\extraspace}}
\newcommand{\ee}{\end{equation}}

\newcommand{\ba}{\begin{eqnarray}
\addtolength{\abovedisplayskip}{\extraspaces}
\addtolength{\belowdisplayskip}{\extraspaces}
\addtolength{\abovedisplayshortskip}{\extraspace}
\addtolength{\belowdisplayshortskip}{\extraspace}}
\newcommand{\ea}{\end{eqnarray}}

\newcommand{\nonu}{\nonumber \\[.5mm]}
\newcommand{\A}{&\!\!\!}

\def\thesection {\S {\arabic{section}}}
\newcommand{\newsection}[1]{
\vspace{7mm} \pagebreak[3] \addtocounter{section}{1}
\setcounter{subsection}{0} \setcounter{footnote}{0}
\begin{center}
{\large {\bf \thesection. #1}}
\end{center}
\nopagebreak
\medskip
\nopagebreak \hspace{3mm}}

\usepackage{multirow}

\begin{document}

\begin{center}
{\bf Energy and momentum  of a spherically symmetric dilaton frame
as regularized by teleparallel gravity}\footnote{ Keywords:
gravitation, teleparallel gravity, energy-momentum,
Weitzenb$\ddot{o}$ck connection, regularized \hspace*{.5cm}
teleparallel gravity}
\end{center}
\begin{center}
{\bf Gamal G.L. Nashed}\footnote{ Mathematics Department, Faculty
of Science, Ain Shams University, Cairo, Egypt.}
\footnote{Egyptian Relativity Group (ERG) URL:
http://www.erg.eg.net}
\end{center}
\bigskip

\centerline{\it Centre for Theoretical Physics, The British
University in Egypt, El-Sherouk City,} \centerline{{\it Misr-Ismalia Desert Road, Postal No. 11837, P.O. Box 43, Egypt.}}

\bigskip
\centerline{ e-mail: \ nashed@bue.edu.eg}

\hspace{2cm} \hspace{2cm}
\\
\\
\\
\\
\\

We calculate  energy and momentum of a spherically symmetric dilaton
frame using the gravitational energy-momentum 3-form within the
tetrad formulation of general relativity (GR).  The
 frame we use is characterized by an arbitrary function $\Upsilon$ with the help of  which
 all the previously found solutions can be reproduced. We show how  the effect
of  inertia  {\it (which is mainly reproduced from  $\Upsilon$)} makes the total energy and momentum always different from
the well  known result when we use the Riemannian connection  ${{\widetilde \Gamma}_\alpha}\vspace{.003cm}^\beta$.
On the other hand, when use is made of the covariant formulation of teleparallel
gravity, which implies to take into account the pure gauge  connection, teleparallel gravity always yields the
physically relevant result for the energy and momentum.

\begin{center}
\newsection{\bf Introduction}
\end{center}
 Our
perspective and understanding of the universe  have changed due to the new discoveries
of the last decades. The discovery of the dark matter and the dark energy
have opened new important questions about the nature of the
matter in Cosmos.
 One of the accepted models to describe the
nature of the dark energy is a scalar field model \cite{CL}.  The dilaton is a  scalar field occurring in the
low energy limit of the  theory where the Einstein action is
supplemented by fields such as the axion, gauge fields and dilaton
coupling in a nontrivial way to the other fields. Exact solutions
for charged dilaton black holes in which the dilaton is coupled to
the Maxwell field have been constructed by many authors. It is
found that the presence of dilaton has  important consequences on
the causal structure and the thermodynamic properties of the black
hole \cite{GM1}$\sim$\cite{HW1}. Thus much interest has been
focused on the study of the dilaton black holes.

Attempts at identifying an energy-momentum density
for gravity has led  to various energy momentum
complexes which are pseudotensors \cite{LL}. Pseudotensors
are not covariant objects i.e., they inherently depend on
the reference frame, and thus  cannot
provide a true  physical local gravitational energy-momentum
density. Hence the pseudotensor approach has been largely abandoned
(cf. \cite{VM}).

It is well known that teleparallel gravity theory allows a separation between
gravitation and inertia \cite{AGP}. Therefore, it turns out possible in
this theory to write down a tensorial
expression for the gravitational energy-momentum density \cite{ALP}.  Computation of the
total energy of Schwarzschild and Kerr spacetimes using a regularized teleparallelism is given in \cite{LOP}.  Obukhov   et al.
\cite{OPR} computed the energy and momentum transported by exact
plane gravitational-wave solutions of Einstein equations using the
teleparallel equivalent of general relativity (TEGR).

The aim of the present work is to calculate the energy and
momentum of a general  spherically symmetric dilaton frame with local Lorentz
transformations containing an arbitrary function $\Upsilon$ which preserve spherical symmetry.
Also we will show how  inertia of energy and momentum are related to a pure gauge (Weitzenb$\ddot{o}$ck) connection ${\Gamma_\alpha}^\beta$\footnote{We
will use the same notation given in Ref. \cite{LOP}.}.
 In \S 2, we use the language of exterior forms to give  an outline
of the teleparallel approach. A brief review is given of the covariant
formalism for the gravitational energy-momentum  which is
described by the pair $( \vartheta^{\alpha},
{\Gamma_\alpha}^\beta)$.  In \S 3, we show
by explicit calculations that due to an inconvenient choice of a
reference frame, the traditional computation of the total
energy and spatial momentum of the spherically symmetric dilaton
solution are unphysical! Using the covariant formalism, we show that the
Weitzenb$\ddot{o}$ck connection acts as a regularizing tool that
separates the inertial contribution and always provides a
physical meaningful result. The final section is devoted for main
results and discussion.\vspace{0.4cm}\\
\centerline{\bf Notation}

We use the Latin indices ${\it i, j, \cdots }$ for local holonomic
spacetime coordinates and the Greek indices $\alpha$, $\beta$,
$\cdots$ label (co)frame components. Particular frame components
are denoted by hats, $\hat{0}$,$\hat{1}$, etc. As usual, the
exterior product is denoted by $\wedge$, while the interior
product of a vector $\xi$ and a p-form $\Psi$ is denoted by $\xi
\rfloor \Psi$. The vector basis dual to the frame 1-forms
$\vartheta^{\alpha}$ is denoted by $e_\alpha$ and they satisfy
$e_\alpha \rfloor \vartheta^{\beta}={\delta}_\alpha^\beta$. Using
local coordinates $x^i$, we have $\vartheta^{\alpha}=h^\alpha_i
dx^i$ and $e_\alpha=h^i_\alpha \partial_i$ where $h^\alpha_i$ and
$h^i_\alpha $ are the covariant and contravariant components of
the tetrad field. We define the volume 4-form by $\eta \stackrel
{\rm def.}{=} \vartheta^{\hat{0}}\wedge \vartheta^{\hat{1}}\wedge
\vartheta^{\hat{2}}\wedge\vartheta^{\hat{3}}.$  Furthermore, with
the help of the interior product, we define \[\eta_\alpha \stackrel
{\rm def.}{=} e_\alpha \rfloor \eta = \ \frac{1}{3!} \
\epsilon_{\alpha \beta \gamma \delta} \ \vartheta^\beta \wedge
\vartheta^\gamma \wedge \vartheta^\delta,\]  where
$\epsilon_{\alpha \beta \gamma \delta}$ is completely antisymmetric
with $\epsilon_{0123}=1$. Furthermore, \[\eta_{\alpha \beta} \stackrel {\rm
def.}{=} e_\beta \rfloor \eta_\alpha =
\frac{1}{2!}\epsilon_{\alpha \beta \gamma \delta} \
\vartheta^\gamma \wedge \vartheta^\delta,\qquad \qquad
\eta_{\alpha \beta \gamma} \stackrel {\rm def.}{=} e_\gamma
\rfloor \eta_{\alpha \beta}= \frac{1}{1!} \epsilon_{\alpha \beta
\gamma \delta} \ \vartheta^\delta,\]  which are bases for 3-, 2-
and 1-forms respectively. Finally, \[\eta_{\alpha \beta \mu \nu}
\stackrel {\rm def.}{=} e_\nu \rfloor \eta_{\alpha \beta \mu}=
e_\nu \rfloor e_\mu \rfloor e_\beta \rfloor e_\alpha \rfloor
\eta,\] is the Levi-Civita tensor density. The $\eta$-forms
satisfy the useful identities: \ba \vartheta^\beta \wedge
\eta_\alpha \A = \A \delta^\beta_\alpha
\eta, \qquad \vartheta^\beta \wedge \eta_{\mu \nu} = \delta^\beta_\nu \eta_\mu-\delta^\beta_\mu \eta_\nu,
\qquad  \vartheta^\beta \wedge \eta_{\alpha \mu \nu}  = \delta^\beta_\alpha \eta_{\mu \nu}+\delta^\beta_\mu
\eta_{\nu \alpha}+\delta^\beta_\nu \eta_{ \alpha \mu}, \nonu
\vartheta^\beta \wedge \eta_{\alpha \gamma \mu \nu}  \A =  \A \delta^\beta_\nu \eta_{\alpha \gamma
\mu}-\delta^\beta_\mu \eta_{\alpha \gamma \nu
}+\delta^\beta_\gamma \eta_{ \alpha \mu \nu}-\delta^\beta_\alpha
\eta_{ \gamma \mu \nu}. \ea

The line element $ds^2 \stackrel {\rm def.}{=} g_{\alpha \beta}
\vartheta^\alpha \bigotimes \vartheta^\beta$ is defined by the
spacetime metric $g_{\alpha \beta}$.

\newsection{Berif review of teleparallel gravity}
Teleparallel geometry can be viewed as a gauge theory of
translation \cite{Hw}$\sim$\cite{Tr}. The coframe
$\vartheta^\alpha$  plays the role of the gauge translational
potential of the gravitational field.  GR can be reformulated as
the teleparallel theory. Geometrically, teleparallel gravity can
be considered as a special case
 of the metric-affine gravity in which
  $\vartheta^\alpha$ and the local Lorentz connection
 are subject to the distant parallelism
constraint ${R_\alpha}^\beta=0$ \cite{OP1}$\sim$\cite{Oy}. In this
geometry the torsion 2-form \be
T^\alpha=D\vartheta^\alpha=d\vartheta^\alpha+{\Gamma_\beta}^\alpha\wedge
\vartheta^\beta=\frac{1}{2}{T_{\mu \nu}}^\alpha \vartheta^\mu
\wedge \vartheta^\nu=\frac{1}{2}{T_{i j}}^\alpha dx^i \wedge
dx^j,\ee arises as the gravitational gauge field strength,
${\Gamma_\alpha}^\beta$ being the Weitzenb$\ddot{o}$ck 1-form
connection, $d$ is the exterior derivative and $D$ is the exterior covariant
derivative. The torsion $T^\alpha$ can be decomposed into three
irreducible pieces \cite{LOP}, the tensor part, the trace and the
axial trace given respectively by  \ba {^ {\tiny{( 1)}}T^\alpha}
\A \stackrel {\rm def.}{=} \A T^\alpha-{^ {\tiny{(
2)}}T^\alpha}-{^ {\tiny{( 3)}}T^\alpha}, \qquad with \nonu
{^  {\tiny{( 2)}}T^\alpha}  \A \stackrel {\rm def.}{=} \A
\frac{1}{3} \vartheta^\alpha\wedge T, \quad where \quad T=
\left(e_\beta \rfloor T^\beta\right), \qquad e_\alpha \rfloor
T={T_{\mu \alpha}}^\mu, \quad vector \ of \  trace \ of \ torsion
\nonu
{^  {\tiny{( 3)}}T^\alpha}  \A \stackrel {\rm def.}{=} \A
\frac{1}{3} e^\alpha\rfloor P, \quad with \quad
P=\left(\vartheta^\beta \wedge T_\beta\right), \quad
e_\alpha\rfloor P=T^{\mu \nu \lambda}\eta_{\mu \nu \lambda
\alpha}, \quad axial \ of \  trace \ of \ torsion.\nonu
\A \A \ea The Lagrangian of the teleparallel equivalent of GR has the
form \be V= -\frac{1}{2\kappa}T^\alpha \wedge ^\ast \left({^
{\tiny{( 1)}}T_\alpha}-2{^  {\tiny{( 2)}}T_\alpha} -\frac{1}{2}{^
{\tiny{( 3)}}T_\alpha} \right), \ee $\kappa=8\pi G/c^3$, $G$ is
the Newton gravitational  constant, $c$ is the speed of light and $\ast$
denotes the Hodge duality in the metric $g_{\alpha \beta}$ which
is assumed to be flat Minkowski metric $g_{\alpha \beta}=o_{\alpha
\beta}=diag(+1,-1,-1,-1)$, that is used to raise and lower local
frame (Greek) indices.

The variation of the total action with respect to the coframe
gives the field equations in the  form  \be
DH_\alpha-E_\alpha=\Sigma_\alpha, \qquad where \qquad  \Sigma_\alpha
\stackrel {\rm def.}{=} \frac{\delta L_{mattter}}{\delta
\vartheta^\alpha},\ee  is the canonical energy-momentum tensor
3-form of matter  which is considered to be the source. In accordance
with the general Lagrange-Noether scheme \cite{Gf, HMM},  one
derives from (4) the translational momentum 2-form and the
canonical energy-momentum 3-form of the gravitational field: \be H_{\alpha} \stackrel {\rm
def.}{=} -\frac{\partial V}{\partial T^\alpha}=\frac{1}{\kappa}
\ast \left({^  {\tiny{( 1)}}T_\alpha}-2{^  {\tiny{( 2)}}T_\alpha}
-\frac{1}{2}{^  {\tiny{( 3)}}T_\alpha} \right), \ee \be E_\alpha
\stackrel {\rm def.}{=} \frac{\partial V}{\partial
\vartheta^\alpha}=e_\alpha \rfloor V+\left(e_\alpha \rfloor
T^\beta \right) \wedge H_\beta. \ee Due to geometric identities
\cite{Oyn}, the Lagrangian (4) can be recast as \be
V=-\frac{1}{2}T^\alpha\wedge H_\alpha.\ee The presence of the
connection field ${\Gamma_\alpha}^\beta$ plays an important
regularizing
role due to the following: \vspace{.3cm}\\
\underline{(i)}: The theory becomes explicitly covariant under
 local Lorentz transformations of the coframe, i.e.,
 the Lagrangian (4) is invariant under the change of
variables \be   \vartheta'^\alpha={\Lambda^\alpha}_\beta
\vartheta^\beta, \qquad
{\Gamma'_\alpha}^\beta={\Lambda^\mu}_\alpha
{\Gamma_\mu}^\nu {(\Lambda^{-1})^\beta}_\nu-{(\Lambda^{-1})^\beta}_\gamma
d{\Lambda^\gamma}_\alpha.\ee
Due to the non-covariant transformation law  of  ${\Gamma_\alpha}^\beta$,
 see Eq. (9), if a connection
vanishes in a given frame, it will not vanish in any other frame related to the first by a
local Lorentz transformation.\vspace{0.5cm}\\
 \underline{(ii)}: ${\Gamma_\alpha}^\beta$ plays an
important role in the teleparallel framework. This role represents
the inertial effects which arise from the choice of the reference
system \cite{ALP}. The  contributions of this inertial object in many
cases lead to unphysical results for the total energy of the
system. Therefore, the role of the teleparallel connection, is to separate the inertial contribution from the truly gravitational
one. Since the teleparallel curvature is zero, the connection is a "pure gauge",
that is
\be {\Gamma_\alpha}^\beta={(\Lambda^{-1})^\beta}_\gamma d {\Lambda^\gamma}_\alpha.\ee
The  Weitzenb$\ddot{o}$ck connection always has the form (10).
The   translational momentum has the form \cite{LOP} \be
\widetilde{H}_\alpha=\frac{1}{2\kappa}{\widetilde{\Gamma}}^{\beta
\gamma}\wedge  \eta_{\alpha \beta \gamma}, \qquad
{\Gamma_\alpha}^\beta \stackrel {\rm def.}{=} {{\widetilde
\Gamma}_\alpha} \vspace{.003cm} ^\beta -{K_\alpha}^\beta,\ee  with ${{\widetilde
\Gamma}_\alpha} \vspace{.003cm} ^\beta$  is the purely Riemannian connection and
$K^{\mu \nu}$ is the contortion 1-form which is related to the
torsion through

\be T^\alpha  =  {K^\alpha}_\beta \wedge
\vartheta^\beta.\ee
\newsection{Total energy of spherically symmetric dilaton spacetime }

In this section, we are going to show how the Weitzenb$\ddot{o}$ck
connection ${\Gamma_\alpha}^\beta$ acts as an inertial object and
contributes to the physical quantities like energy, momentum etc.,
when it is trivial. On the other hand,  when this connection is non
trivial, we  show how it separates the inertia from the other physical
quantities. We will show this by studying a spherically symmetric
spacetime which contains an arbitrary function $\Upsilon$ which preserve spherical symmetry and
reproduce all the previous solutions.  {\it This study is carried out for the spherically
symmetric cases only.}

Using the spherical local coordinates $(t,r,\theta, \phi)$,
the spherically symmetric dilaton  is described by the coframe components: \be
{{\vartheta^{{}^{{}^{\!\!\!\!\scriptstyle{S}}}}}{_{}{_{}{_{}}}}^{\alpha}}
={{\Lambda_1}^\alpha}_\gamma
{{\Lambda_2 }^\gamma}_\delta \vartheta^{\delta}, \ee
where the coframe $\vartheta^{\delta}$ has the form \ba
\vartheta^{\hat{0}}\A=\A\alpha^{-1}cdt,\quad\vartheta^{\hat{1}}=\alpha
dr, \quad \vartheta^{\hat{2}}=r \ \beta \ d\theta, \quad
\vartheta^{\hat{3}}=r \ \beta \ \sin\theta \  d\phi, \qquad where\nonu
\A \A  \alpha= \left(1-\frac{2m}{r}\right)^{-\frac{1}{2}}, \qquad  \qquad \beta=\sqrt{1-\frac{q^2 e^{-2\phi_0}}{mr}}, \ea where $m$, $q$  and  $\phi_0$ are the mass,  the charge
and  the asymptotic value of the dilaton, respectively.   The matrices ${{\Lambda_1}^\alpha}_\gamma$ and
${{\Lambda_2 }^\gamma}_\delta $ are the  local Lorentz
transformations that are defined respectively as \be
{{\Lambda_1}^\alpha}_\gamma  \stackrel {\rm def.}{=}
\left( \matrix{ 1 &  0 & 0 & 0 \vspace{3mm} \cr  0  &  \sin\theta
\cos\phi &  \cos\theta \cos\phi & - \sin\phi \vspace{3mm} \cr 0  &
\sin \theta \sin \phi& \cos\theta \sin\phi & \cos\phi \vspace{3mm}
\cr 0  & \cos\theta & -\sin\theta  & 0 \cr }\right)\; ,\ee \ba
{{\Lambda_2 }^\gamma}_\delta \A \stackrel {\rm def.}
{=} \A \left( \matrix{ \beta_1 & \beta_2 \sin\theta \cos\phi & \beta_2
\sin\theta \sin\phi &  \beta_2 \cos\theta \vspace{3mm} \cr - \beta_2
\sin\theta \cos\phi & 1+\beta_3\sin^2\theta
\cos^2\phi &\beta_3\sin^2\theta \sin\phi \cos\phi
&\beta_3\sin\theta \cos\theta \cos\phi \vspace{3mm}
\cr - \beta_2 \sin\theta \sin\phi &\beta_3 \sin^2\theta
\sin\phi \cos\phi &1+\beta_3\sin^2\theta \sin^2\phi
&\beta_3 \sin\theta \cos\theta \sin\phi \vspace{3mm}
\cr - \beta_2 \cos\theta &\beta_3\sin\theta \cos\theta
\cos\phi &\beta_3\sin\theta \cos\theta \sin\phi
&1+\beta_3\cos^2\theta \cr}\right), \nonu
\A \A
 with \qquad   \beta_1=1+\sqrt{e^{2\Upsilon}+1}-\sqrt{2}, \nonu
 \A \A  \beta_2=\sqrt{3+2\sqrt{e^{2\Upsilon}+1}\left(1-\sqrt{2}\right)+e^{2\Upsilon}-2\sqrt{2}}, \qquad \beta_3=\beta_1-1,
 \ea where $\Upsilon=\Upsilon(r)$ is an arbitrary function. It can be shown that  $\Upsilon(r)$ can reproduce the previous
 arbitrary function $H(r)$ studied in ([36], Eq. (16)) through the relation\footnote{When $H(r)=0$, Eq. (17) gives $\Upsilon=0$ and  Eq. (16) gives the  identity matrix which will be identical with Eq. (3$\cdot$7) in \cite{Nprd3}, Eq. (16) in Ref. \cite{Nprd1} and Eq. (13) in \cite{Nprd} when we used  the proper Lorentz transformation. }
 \be \Upsilon(r)=\ln \sqrt{2\sqrt{H^2(r)+1}(\sqrt{2}-1)+H^2(r)+3-2\sqrt{2}} \ \ .\ee
  The metric tensor $g_{i j}\stackrel {\rm def.}{=}  o_{\mu \nu} {h^\mu}_i {h^\nu}_j$ associated with the tetrad field (13) has the form
\[
ds^2= \frac{1}{\alpha^2}dt^2
-\alpha^2 dr^2 -r^2\beta^2 d\Omega^2, \quad with \quad  d\Omega^2=d\theta^2+\sin^2\theta d\phi^2,
 \] which is dilaton spacetime derived in \cite{GHS}.

  From Eq. (17) all the previous spherically symmetric
 solution can be obtained \cite{Nprd}.
  If we take tetrad (13), as well as
the trivial Weitzenb$\ddot{o}$ck connection
${\Gamma^\alpha}_\beta=0$ and substitute into (11),  we finally get
\ba \widetilde{H}_{\hat{0}}\A=\A-\frac{\sin\theta  \beta r}{8\pi}
\Biggl\{\left(\cos\phi\sin\theta\cos\theta\sin\phi+\cos^2\phi\sin^2\theta+\cos\theta\sin\phi+2\frac{\beta' r+\beta}{\alpha}-1\right)\beta_3\nonu
\A \A-2\left(1-\frac{\beta' r+\beta}{\alpha}\right)\Biggr\}d\theta\wedge d\phi.\ea If we
compute the total energy at a fixed time in the 3-space with a
spatial 2-dimensional boundary surface $\partial S =\{r = R,
\theta,\phi\}$,  we obtain \be \widetilde{E} =\int_{\partial S}
\widetilde{H}_{\hat{0}} =\frac{2R \beta}{3}\left(\beta_1\left[1-\frac{3[\beta'+\beta]}{\alpha}\right]
\right).\ee

Using Eq. (19) we discuss the following cases:\vspace{.5cm}\\
(i) When $\Upsilon(R)=0$, then Eq. (16) gives $\beta_1=1$, $\beta_2=0$ and $\beta_3=0$.  In this case the energy takes the  form up to $O\left(\frac{1}{R}\right)$
\be \widetilde{E
}\cong m-\frac{q^2e^{-2\phi_0}-m^2}{2R}-\frac{q^4e^{-4\phi_0}}{8m^2R}+O\left(\frac{1}{R^2}\right),\ee which is  consistent with the previous result  (\cite{Nprd2}, Eq (45)). In this case the  local Lorentz
transformation given by Eq. (16) will be identical with the
Kronecker delta, i.e.,  ${\delta}^\alpha_\beta=diag(+1,+1,+1,+1)$. \vspace{.5cm}\\
(ii) When  $\Upsilon(R)\cong \displaystyle\frac{c_1}{\sqrt{R}}$ we get
\be \widetilde{E}= -\frac{{c_1\sqrt{2R}}}{3}+m-\frac{{c_1}^2}{2\sqrt{2}}+\frac{c_1 q^2e^{-2\phi_0}+3c_1}{3m\sqrt{2R}}-\frac{q^2e^{-2\phi_0}-m^2}{2R}-\frac{q^4e^{-4\phi_0}}{8m^2R}+O\left(\frac{1}{R^{3/2}}\right),\ee
which is a divergent one. It is clear from Eq. (21)  how the inertia $c_1$ contributes the physical quantities.\vspace{.5cm}\\
(iii) When  $\Upsilon(R)\cong \displaystyle\frac{c_1}{{R}}$, then the energy takes the form
\be \widetilde{E}= m-\frac{{c_1\sqrt{2}}}{3}+\frac{2c_1 q^2e^{-2\phi_0}+3{c_1}^2-3m\sqrt{2} q^2e^{-2\phi_0}+3m^3\sqrt{2} }{6m\sqrt{2R}}+O\left(\frac{1}{R}\right),\ee
which is not divergent but not consistence with the previous results (\cite{Nprd2}, Eq. (72)). If we continue in this manner, i.e.,  $\Upsilon(R)\cong \displaystyle\frac{c_1}{{R^{3/2}}}$ or  $\Upsilon(R)\cong \displaystyle\frac{c_1}{{R^2}}$,   we can show that
the inertia will continue in its contribution  to the physical quantities up to order $O\left(\frac{1}{R^2}\right)$. \vspace{.5cm}\\
(iv) When  $\Upsilon(R)\cong \displaystyle\frac{c_1}{{R^{5/2}}}$
the form of energy will be the same as given by Eq. (20).

 To overcome the above  problems (divergent or contribution of inertia to physical quantities) we are
going to use the regularization framework which is based on the
covariance property, i.e., we will take into account the
Weitzenb$\ddot{o}$ck connection ${\Gamma^\alpha}_\beta$ given by
 Eq. (10) in which  ${{\Lambda}^\alpha}_\beta={{\Lambda_1}^\alpha}_\gamma
 {({{\Lambda_2}^{-1}})^\gamma}_\delta {({{\Lambda_1}^{-1}})^\delta}_\beta$ \cite{LOP}.

Using the regularization framework   and calculating the necessary components, we finally get
the superpotential \ba H_{\hat{0}}\A =\A
\frac{\beta r \sin\theta}{4\pi}\Biggl[\left\{\sqrt{e^{2\Upsilon}+1}(\sqrt{2}-1)
-e^{2\Upsilon}-1\right\}(\beta' r+\beta-\alpha)
\Biggr] d\theta\wedge d\phi.\ea The total
energy of (23) thus has the form \be E=\int_{\partial S}
H_{\hat{0}}= R\beta\Biggl[\left\{\sqrt{e^{2\Upsilon}+1}(\sqrt{2}-1)
-e^{2\Upsilon}-1\right\}(\beta' r+\beta-\alpha)
\Biggr].\ee
From Eq. (24) we discuss the following:\vspace{.5cm}\\
 If  $\Upsilon(R)=0$ or  $\Upsilon(R)=\frac{c_1}{\sqrt{R}}$ or  $\Upsilon(R)=\frac{c_1}{R}$  etc., the value of the energy will have the form of Eq. (20) which is consistence with the previous results (\cite{Nprd2} Eq. (45))\footnote{Quadratic terms like $c_1$m are neglected in this approximation.}.

The non vanishing
components  needed to calculate the spatial momentum
$\widetilde{H}_{\hat{\alpha}}=H_{\hat{\alpha}}, \\
\hat{\alpha}=1,2,3$ have the form \ba
\widetilde{H}_{\hat{1}}=H_{\hat{1}} \A= \A
\frac{\beta\beta_2r(\beta'r+\beta-\alpha)\sin\theta\left[\sin\phi
\cos\phi\{1-\sin\theta\cos\theta\}-\cos^2\phi\sin^2\phi\right]}{4\alpha\pi}
d\theta\wedge d\phi,\nonu
\widetilde{H}_{\hat{2}}=H_{\hat{2}} \A= \A
\frac{\beta\beta_2 r(\beta' r+\beta-\alpha)\sin\theta\left[\cos\theta
\cos\phi(\sin\theta\cos\phi-1)-\cos\phi\sin\phi\sin^2\theta-\cos\theta\sin\theta\right]}{4\alpha\pi}
d\theta\wedge d\phi, \nonu
\widetilde{H}_{\hat{3}}=H_{\hat{3}} \A= \A
\frac{\beta\beta_2r(\beta' r+\beta-\alpha)\sin^2\theta(\cos\theta
\cos\phi-\sin\theta\sin\phi)}{4\alpha\pi} d\theta\wedge d\phi.
\ea Using Eqs. (25) in Eq. (11), we finally get the spatial momentum in the
form \be P_{\hat{\alpha}}=\int_{\partial S} H_{\hat{\alpha}}=0,\qquad \hat{\alpha}=1,2,3.\ee

\newsection{Main results and Discussion}
The main results of this paper are the following:\vspace{0.4cm}\\
$\bullet$  Of new local Lorentz transformation with an arbitrary function $\Upsilon(r)$ which maintain spherical symmetry is given by Eq. (16).
The relation between this transformation and
the transformation studied in \cite{Nprd} is given through Eq. (17). \vspace{0.3cm}\\
$\bullet$  The frame we have studied creates the same spacetime as the one derived in \cite{GHS}. This spacetime is characterize by the
gravitational mass $m$, the charge  $q$ and the asymptotic value of the dilaton $\phi_0$.   \vspace{0.3cm}\\
$\bullet$ We have calculated the energy of the frame (13) by using two procedures: \vspace{0.3cm}\\
(i) In the first procedure, we have taken the Riemannian connection only and we have shown that the energy may be divergent or not be of the well know form.
 We  have explained how the form of energy depends on the asymptotic value of the arbitrary function $\Upsilon(R)$.
  When the arbitrary function  $\Upsilon(R)$
 is asymptotically smaller than $\frac{1}{R^{5/2}}$, then the form of energy may be divergent or not in agreement with the previous result (\cite{Nprd2}, Eq. (45)).
 On the other hand, if  $\Upsilon(R)$
 is asymptotically greater or equal to  $\frac{1}{R^{5/2}}$,  the form of energy will be in  agreement with the previous result \cite{Nprd}.  \vspace{0.3cm}\\
 (ii) When use is made of the covariant formulation of teleparallel
gravity, which implies to take into account the pure gauge (or
 Weitzenb$\ddot{o}$ck) connection given in Eq. (10). Teleparallel gravity always yields the
physically relevant result for the energy and momentum.\vspace{0.3cm}\\
$\bullet$ The second argument in the previous item is known as the regularization of the teleparallelism in which the
covariant teleparallel approach always yields the physically correct result.
 \vspace{0.3cm}\\

\bigskip
\bigskip
\centerline{\Large{\bf Acknowledgements}}

The author would like to thank Professor M. I.  Wanas; Cairo
University, for his stimulating discussions.\\

A comparison between the results  of the arbitrary function $\Upsilon(r)$ which
reproduces  the dilaton spacetime and its energy
using different translational momenta  are given in the following
table.

\newpage
\begin{center}
Table 1.  Comparison between the results of the arbitrary
function $\Upsilon(r)$ which keeps  spherical symmetry and the other
constants which create dilaton spacetime and
its energy using different translational momenta
\end{center}
\begin{center}

\begin{tabular}{|c|l|c|l|l|}
\hline
Spacetime & Arbitrary & The constant  & Translational Momentum & Energy \& References\\
&function $\Upsilon(r)$&$c_1$ &&\\ \hline
 &  \multirow{5}{*} &  & Equation (11) with trivial& \\
& $\Upsilon(r)=0$, $q=0$  &  $c_1=0$ & Weitzenb$\ddot{o}$ck connection,&$E=m$, Eq. (20),  \\
&in Eq. (13)&& i.e., ${\Gamma^\alpha}_\beta = 0$ & Ref. [36] \\\cline{2-5}

&&&Equation (11) with trivial&\\
Schwarzschild  & $\Upsilon(r)\neq0$, $q=0$  &  $c_1=m$  & Weitzenb$\ddot{o}$ck connection,&$E\neq m$, Eq. (21), \\
  &in Eq. (13)&& i.e., ${\Gamma^\alpha}_\beta = 0$ &Ref. [36]  \\\cline{2-5}
  & $\Upsilon(r)\neq 0$, $q=0$&  & Equation (11) with trivial & \\   &  in Eq. (13)&  $c_1\neq0$  &
 Weitzenb$\ddot{o}$ck connection, &$E\neq m$,  \\ &&&  i.e., ${\Gamma^\alpha}_\beta = 0$ & Eq. (21) \\\cline{2-5}

 & $\Upsilon(r)\neq 0$, $q=0$&  & Equation (11) with non & \\   &  in Eq. (13)&  $c_1\neq0$  &
 trivial Weitzenb$\ddot{o}$ck connection, &$E=m$,  \\ &&&  i.e., ${\Gamma^\alpha}_\beta \neq 0$ & Eq. (24)
\\
 \hline
& \multirow{5}{*}&&Equation (11) with trivial & \\
&  $\Upsilon(r)=0$&  $c_1=0$ & Weitzenb$\ddot{o}$ck connection, &$E= m-\frac{q^2}{2r}$,
\\ &&$\phi_0=0$&  i.e., ${\Gamma^\alpha}_\beta = 0$ & Eq. (20), Ref. [36]
\\
 \cline{2-5}
Reissner-  & &  &Equation (11) with trivial & \\Nordstr$\ddot{o}$m &
$\Upsilon(r) \neq 0$&  $c_1=m$  & Weitzenb$\ddot{o}$ck connection,&$E\neq
m-\frac{q^2}{2r}$,  \\ &&$\phi_0=0$& i.e., ${\Gamma^\alpha}_\beta = 0$  & Ref. [36]
\\  \cline{2-5}
& &  & Equation (11) with  trivial & \\   &  $\Upsilon(r) \neq 0$&
$c_1\neq0$ &
 Weitzenb$\ddot{o}$ck connection, &$E\neq m-\frac{q^2}{2r}$,  \\ && $\phi_0=0$ &  i.e., ${\Gamma^\alpha}_\beta = 0$ & Eq. (20)
 \\  \cline{2-5}
& &  & Equation (11) with non trivial & \\   &  $\Upsilon(r) \neq 0$&
$c_1\neq0$ &
 Weitzenb$\ddot{o}$ck connection, &$E= m-\frac{q^2}{2r}$,  \\ && $\phi_0=0$ &  i.e., ${\Gamma^\alpha}_\beta \neq 0$ & Eq. (24)
 \\
\hline
& \multirow{5}{*}&&Equation (11) with trivial & \\
&  $\Upsilon(r)=0$&  $c_1=0$ & Weitzenb$\ddot{o}$ck connection, &$E=m-\frac{q^2e^{-2\phi_0}-m^2}{2R}$,
\\ &&$\phi_0\neq0$&  i.e., ${\Gamma^\alpha}_\beta = 0$ & Ref. [36]
\\
 \cline{2-5}
Dilaton  & &  &Equation (11) with trivial& \\spacetime &
$\Upsilon(r) \neq 0$&  $c_1=m$  & Weitzenb$\ddot{o}$ck connection,&$E\neq
m-\frac{q^2e^{-2\phi_0}-m^2}{2R}$,  \\ &&$\phi_0\neq0$& i.e., ${\Gamma^\alpha}_\beta = 0$  & Ref. [36]
\\  \cline{2-5}
& &  & Equation (11) with  trivial & \\   &  $\Upsilon(r) \neq 0$&
$c_1\neq0$ &
 Weitzenb$\ddot{o}$ck connection, &$E\neq m-\frac{q^2e^{-2\phi_0}-m^2}{2R}$,  \\ && $\phi_0\neq0$ &  i.e., ${\Gamma^\alpha}_\beta = 0$ & Eq. (21)
 \\  \cline{2-5}
& &  & Equation (11) with non trivial & \\   &  $\Upsilon(r) \neq 0$&
$c_1\neq0$ &
 Weitzenb$\ddot{o}$ck connection, &$E=m-\frac{q^2e^{-2\phi_0}-m^2}{2R}$,  \\ && $\phi_0\neq0$ &  i.e., ${\Gamma^\alpha}_\beta \neq 0$ & Eq. (24)
 \\
\hline
\end{tabular}
\end{center}

\end{document}